\documentclass[twocolumn]{aastex631}

\usepackage{amsmath}
\usepackage{xspace}

\newcommand{\unit}[1]{\,\mathrm{#1}}
\renewcommand{\vec}[1]{\boldsymbol{#1}}
\newcommand{\figscale}[0]{0.8}

\newcommand{\enzo}[0]{ENZO\xspace}
\newcommand{\HESS}[0]{H.E.S.S.\xspace}
\newcommand{\Fermi}[0]{\textit{Fermi} LAT\xspace}
\newcommand{\ES}[0]{1ES 0229+200\xspace}

\begin{document}

\title{Constraining the astrophysical origin of intergalactic
magnetic fields}

\author{J. Tjemsland}
\affiliation{Department of Physics, Norwegian University of Science and Technology \\
Høgskoleringen 5, 7491 Trondheim, Norway}

\author{M. Meyer}
\affiliation{CP3-Origins, University of Southern Denmark \\
Campusvej 55, 5230 Odense M, Denmark}

\author{F. Vazza}
\affiliation{Dipartimento di Fisica e Astronomia, Universitá di Bologna \\
Via Gobetti 93/2, 40129 Bologna, Italy}
\affiliation{INAF-Istitituto di Radio Astronomia \\
Via Gobetti 101, 40129 Bologna, Italy}
\affiliation{Hamburger Sternwarte, Universität Hamburg \\
Gojenbergsweg 112, 41029 Hamburg, Germany}

\begin{abstract}
High-energy photons can produce electron-positron pairs upon interacting with the
extragalactic background
light (EBL). These pairs will in turn be deflected by the intergalactic magnetic field (IGMF),
before possibly up-scattering photons of the cosmic microwave background (CMB),
thereby initiating an electromagnetic cascade.
The non-observation of an excess of GeV photons and an extended halo around
individual blazars due to this electromagnetic cascade can be used to constrain the
properties of the IGMF.
In this work, we use publicly available data of \ES by \Fermi and \HESS
to constrain cosmological MHD simulations of various magnetogenesis
scenarios, and find that all models without a strong space-filling primordial
component or over-optimistic dynamo amplifications can be excluded at the 95\%
confidence level. In fact,  we find that the fraction of space filled by a strong IGMF
has to be at least $f\gtrsim 0.67$, thus excluding most astrophysical production scenarios.
Moreover, we set lower limits
$B_0>5.1\times 10^{-15}$ G ($B_0>1.0\times 10^{-14}$ G) for a space-filling
primordial IGMF
for a blazar activity time of $\Delta t = 10^4$ yr ($\Delta t = 10^7$ yr).
\end{abstract}


\section{Introduction}

The main hypothesis for the production of magnetic fields in astrophysical
structures, from stars up to galaxy clusters, is that the magnetic fields are 
produced by amplifications of pre-existing magnetic fields by various dynamo 
effects or during gravitational collapse~\citep{Durrer:2013pga}. In order to explain
the magnetic fields observed in galaxy clusters, one therefore usually presumes that
the intergalactic magnetic field (IGMF) has a seed field that was present before
galaxies were formed. This seed field is known as the primordial magnetic field, and will still
be present in cosmic voids around its original strength,
diluted only due to the expansion of the Universe and affected by (inverse) energy cascades.
The production mechanism of the primordial field remains, however, still unknown.
Thus, by constraining or detecting the IGMF, one can learn about unknown
processes in the early Universe, e.g., deduce whether the
IGMF was produced during inflation or during phase
transitions~\citep{Durrer:2013pga}.
Unfortunately, the primordial field will be weak, making it difficult to detect. Indeed, 
Faraday rotation measurements provide currently an upper limit on the order
of $10^{-9}\unit{G}$~\citep{Pshirkov:2015tua}.
Alternatively, the IGMF could have an astrophysical origin (see below).

The IGMF can be probed by observing gamma-rays from blazars~\citep{Neronov:2009gh}:
High-energy photons can produce electron-positron pairs upon interacting with the
extragalactic background
light (EBL). The pairs will in turn be deflected by the IGMF, before they may up-scatter 
cosmic microwave background (CMB) photons via the inverse Compton scattering, thereby
initiating an electromagnetic cascade.
The deflection of the electrons and positrons leads to a time delayed arrival of
cascade photons compared to the primary photons~\citep{Plaga:1995ins}.
In addition, the cascade photons will
reach Earth slightly off-angle compared to the line-of-sight, leading to an extended
halo around the primary point-source~\citep{Aharonian:1993vz}.
The size of the halo and the time-delay depend on
the strength of the IGMF: A strong IGMF implies a long time-delay and a wide
cascade halo with low brightness.

The non-observation of an excess of GeV photons of individual blazars using \Fermi,
first suggested by \citet{dAvezac:2007xri},
has been used to set a lower limit on the IGMF
of $B\gtrsim 10^{-15}\text{--}10^{-16}\unit{G}$~%
\citep[e.g.][]{Tavecchio:2010mk,Neronov:2010gir,Dolag:2010ni,Podlesnyi:2022ydu}. Likewise, using the
non-observation of cascade halos with imaging air Cherenkov telescopes have been used to
rule out $B\sim 10^{-16}$ using
\HESS~\citep{HESS:2014kkl} and $B\sim 10^{-14}$ using
VERITAS~\citep{VERITAS:2017gkr}. The limits weaken, however, considerably for low blazar activity
times~\citep{Dolag:2010ni,Dermer:2010mm} while becoming stronger
for coherence lengths $l_\mathrm{B}\lesssim 0.1\unit{Mpc}$.
A recent and comprehensive review of existing limits on the IGMF is given by
\citet{AlvesBatista:2021sln}.
Currently, the most robust lower limit from blazar observations is 
$B>1.8\times10^{-14}\unit{G}$ ($B>3.9\times 10^{-14}\unit{G}$) for an
activity time of $10^4\unit{yr}$ ($10^7\unit{yr}$), set by a combined analysis of \Fermi and
\HESS blazar observations~\citep{HESS:2023zwb}.

Even though current limits focus on a space-filling primordial field, it is likely
that the IGMF has a complicated morphology induced by dynamo amplifications around 
galaxies and galaxy clusters, as well as outflows from magnetized galaxies.
Indeed, an alternative production mechanism of the IGMF is that the magnetic fields
are produced at late times by large-scale outflows from active magnetized 
galaxies~\citep{Durrer:2013pga}. This will allow for a miniscule magnetic field in
the cosmic voids. Nevertheless, the lower limits from the blazar observations will
in this case become tighter, and can be used to limit the filling fraction, $f$,
of the IGMF~\citep{Dolag:2010ni}.

In this work, we make use of the recent \Fermi and \HESS observations
of \ES by \citet{HESS:2023zwb} to constrain the astrophysical origin
of the IGMF.
By comparing observations to Monte Carlo simulations performed using
the photon propagation program
\texttt{ELMAG}~\citep{Blytt:2019xad,Kachelriess:2011bi},
we find a lower limit 
$B_0>5.1\times 10^{-15}$ G ($B_0>1.0\times 10^{-14}$ G) assuming a blazar
activity time of $\Delta t = 10^4$ yr ($\Delta t = 10^7$ yr),
comparable to the results by \citet{HESS:2023zwb}.
The main emphasis in this work is, however, on the astrophysical origin
of the IGMF. By using a top-hat distribution as a proxy for an astrophysical
magnetic field, we find that the IGMF must have a filling
fraction of at least $f>0.67$, which excludes most astrophysical
production scenarios.
Furthermore, we perform the analysis on numerical
models of magnetogenesis produced using cosmological MHD
simulations~\citep{Vazza:2017qge,Gheller:2019wlf}.
We find that none, even in the excessive galaxy formation feedback models, can
magnetize the Universe in a way that is compatible with blazar data. 
Only models with a strong primordial magnetic field or over-optimistic dynamo
amplifications cannot be excluded at the 95\% confidence level (CL).

\section{Observations of \ES}

The blazar \ES is a high frequency peaked BL Lac object located
at a redshift $z=0.14$, with a hard spectrum extending to
10 TeV~\citep{Kaufmann:2011az}. 
This makes it an ideal source to search for cascade photons and
extended halos, and it has therefore been the primary source for setting
bounds on the IGMF for a long time~\citep[e.g.][]{Dolag:2010ni,HESS:2014kkl}.
Recently, the observed \Fermi and \HESS spectra of \ES were made publicly
available~\citep{HESS:2023zwb}\footnote{
  The data from \citet{HESS:2023zwb} used in this work can be accessed
    via \citet{manuel_meyer_2023_8014311} at \dataset[10.5281/zenodo.8014311]{\doi{10.5281/zenodo.8014311}}.
}, which we make use of in the present work.

The \Fermi observations comprise 11.5\,yr of data leading to a detection of \ES at 14.2\,$\sigma$. 
The considered \HESS data consists of 144\,hr (dead-time corrected) taken with the small 12\,m diameter H.E.S.S. telescopes until 2016. The observations yield a source significance of 16.5\,$\sigma$. 
Both \Fermi and \HESS spectral can be well described with a simple power law, $\phi(E) = N_0 (E/E_0)^{-\Gamma}$, with spectral index $\Gamma = 1.77 \pm 0.09$ in the \Fermi energy band and $\Gamma = 2.81 \pm 0.11$ in the \HESS energy band, respectively. Absorption of gamma rays on the EBL can account for the entire spectral softening. 
Further details on the analysis can be found in~\citet{HESS:2023zwb}.

\section{IGMF models}
\label{sec:igmf}

\begin{figure}
    \centering
    \includegraphics[scale=\figscale]{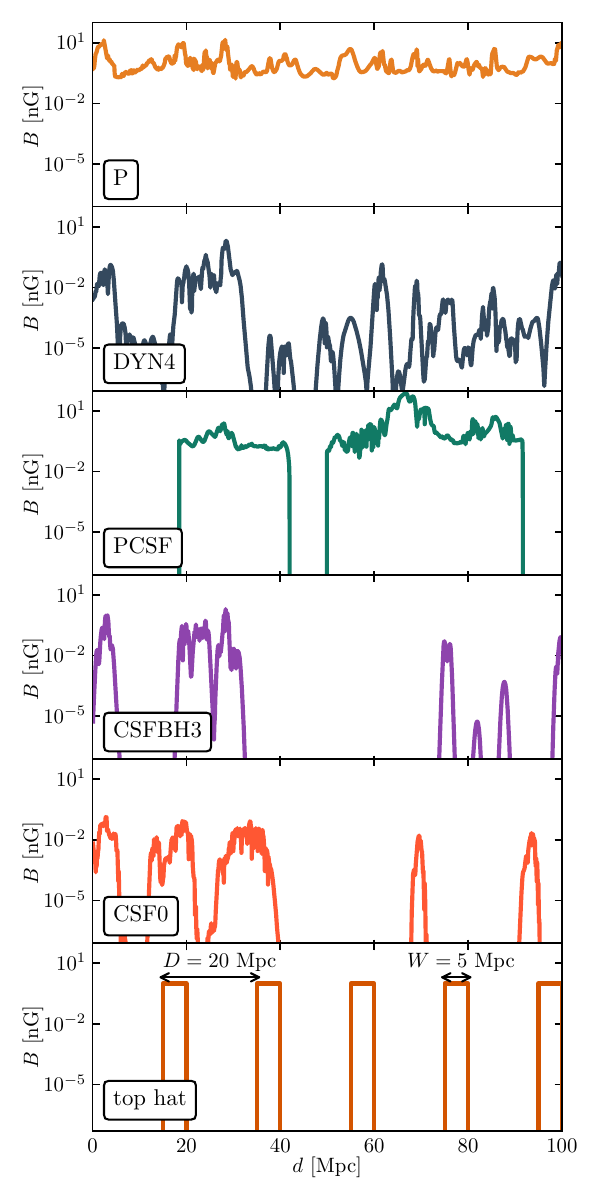}
    \caption{Visualization of the magnetic field strength of some of the
    magnetic field models considered: P, DYN4, PCSF, CSFBH3, CSF0 and a top-hat
    distribution. The models are described in more detail in the main
    text.
    }
    \label{fig:bfield}
\end{figure}

In this work, we make use of a selection of theoretical models of magnetogenesis
produced using cosmological MHD simulations with the ENZO\footnote{\url{http://enzo-project.org}} code~\citep{ENZO:2013hhu},
known as the ``Chronos++ suite''~\citep{Vazza:2017qge,Vazza:2021vwy}\footnote{
    See \url{http://cosmosimfrazza.eu/the\_magnetic\_cosmic\_web} for more information.
}.
These explore numerous plausible scenarios
for radiative gas cooling, star and AGN feedback and dynamo amplification,
starting from the (non-radiative)
evolution of large-scale magnetic fields with initial seed value of
$B_0=10^{-9} \unit{G}$, $B_0=10^{-11}\unit{G}$ or $B_0=10^{-18}\unit{G}$ (comoving).
All simulations start at redshift $z=38$ and 
evolve at the constant comoving spatial resolution of $83.3\unit{kpc}$ per cell. More
specific details of the tested astrophysical models are given in Appendix~\ref{app:B_details}.
Portions of a line-of-sight magnetic field for five of the considered
magnetogenesis models are visualized in Fig.~\ref{fig:bfield}:
A space-filling primordial field with seed value $B_0=1\unit{nG}$ (P),
a weak initial primordial field significantly
enhanced by efficient dynamo amplification (DYN4),
a model with extreme magnetization from astrophysical feedback
(i.e. even beyond the likely feedback power of real low mass galaxies) (PCSF),
and two with a weak initial primordial field and stellar feedback (CSFBH3 and CSF0).
The models PCSF and DYN4 are, as we will see,
in tension with current gamma-ray data, but cannot be excluded at the 95 \% CL.
Meanwhile, P cannot be excluded due to its high seed value, and CSFBH3 and CSF0 will
be excluded due to their small filling fraction.

In addition to the MHD simulations, we consider some common
simplified magnetic field models in order to draw some general conclusions
about the allowed properties of the IGMF.
For simplicity, we model the turbulent field as a
simple domain-like magnetic field, where the field is split into
homogeneous patches of size equal to the coherence length
$l_\mathrm{B}$. The magnetic field strength in each patch has a constant strength
$B_0$, but a random direction. Since the
cascade predictions remain independent of the coherence length for
$l_\mathrm{B}\gtrsim 0.1\unit{Mpc}$, we fix for simplicity
$l_\mathrm{B}=1\unit{Mpc}$ throughout this paper.
As a proxy for an IGMF of a purely astrophysical origin, we consider a simple
top-hat magnetic field, as shown in the bottom panel of Fig.~\ref{fig:bfield}. The
magnetic field strength in the voids is $B_0=0$, while the magnetic
field in the filaments is assumed to be sufficiently strong that the
electrons and positrons in the electromagnetic cascade are completely
isotropized. This 
model has then only two parameters: the width of filaments, $W$, and the filling
fraction, $f$. In the particular example in Fig.~\ref{fig:bfield},
the filling fraction is $f=W/D=0.25$, where $D$ is the distance between the onset of two high magnetic field regions.

\section{Cascade predictions}

We simulate the electromagnetic cascade using the \texttt{ELMAG 3.02} Monte Carlo
code~\citep{Kachelriess:2011bi,Blytt:2019xad}. For each magnetic field
model, we inject photons with energies logarithmically distributed
between 1 GeV and 100 TeV, and the results are stored in a multidimensional
histogram depending on the primary energy of the initial photon, the secondary
energy of the arriving photons at Earth, the time delay and
the arrival angle.
In this way, we can as a second step easily perform a fit to the experimental data by
re-scaling the results by the source spectrum, as done in \citet{HESS:2023zwb}. 
We consider the photons arriving within the 95\% containment radius of the point-spread function (PSF) of \Fermi around
\ES. For simplicity, we follow \citet{HESS:2023zwb} by fixing the
jet opening angle $\theta_\mathrm{jet}=6^\circ$ and assume that the jet is aligned with the line-of-sight,
$\theta_\mathrm{obs}=0$.
Furthermore, we use the EBL model by Dominguez et al.~\citep{Dominguez:2010bv}, as implemented
in \texttt{ELMAG}.
More details about the simulations
and the fits are given in Appendices~\ref{app:simulation} and~\ref{app:details_on_the_fit}.

\begin{figure*}
    \centering
    \includegraphics[scale=\figscale]{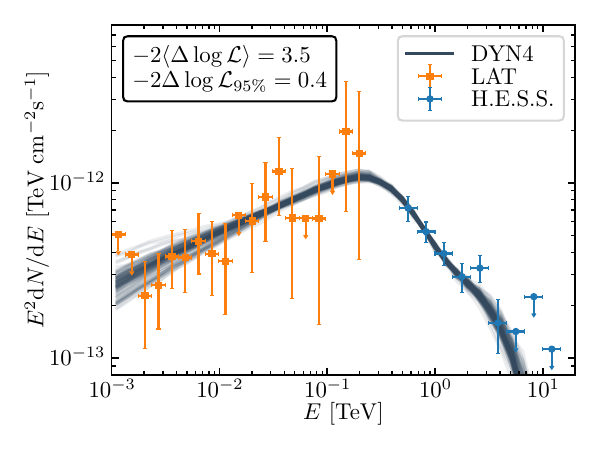}
    \includegraphics[scale=\figscale]{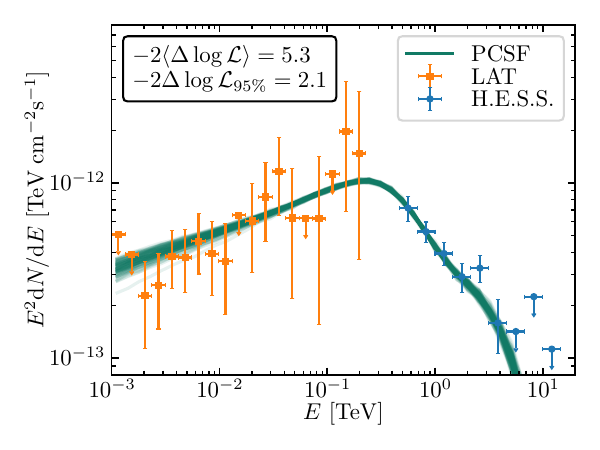}
    \caption{Example plot of the cascade prediction for the MHD models DYN4 (left) and PCSF (right).
    Each of the 100 lines correspond to a random line-of-sight from the simulation volume of
    the MHD simulation.
    The source spectrum is adjusted so that the final spectrum minimizes the maximum likelihood estimate
    of the experimental measurements by Fermi LAT (squares) and H.E.S.S (circles), as explained in
    Sec.~\ref{sec:analysis_and_results}.
    }
    \label{fig:cascade_example}
\end{figure*}

In contrast to the domain-like magnetic field, the effective filling
fraction in the MHD simulations will depend on the chosen line-of-sight in the simulation volume. 
We therefore have to consider a sufficiently large amount of possible
line-of-sights through the MHD simulation volume. 
To limit the necessary computational time, we have extracted
100 line-of-sight magnetic fields for each of the considered magnetic field
models that we use in the analysis. 
Note that periodic boundary conditions are
used since the size of the box in the MHD simulations, $D=83.3\unit{Mpc}$,
is shorter than the distance to \ES, $d\approx 591\unit{Mpc}$.
Nevertheless, this will lead to a conservative
estimation, since the cosmic
variance of a larger simulation volume would lead to a smaller variation
in the final cascade spectrum.
Two examples of the cascade prediction
are shown in Fig.~\ref{fig:cascade_example} for visualization.

\section{Analysis and results}
\label{sec:analysis_and_results}

In this work, we use the publicly available spectra from~\citet{HESS:2023zwb}
to constrain properties of the IGMF. 
As in \citet{HESS:2023zwb}, we model the point source spectrum as
\begin{equation}
    \phi_\mathrm{src} = N_0\left(\frac{E}{E_0}\right)^{-\Gamma}
      \exp\left(-\frac{E}{E_\mathrm{cut}}\right),
\end{equation}
where $N_0$ is the normalization, $E_0\equiv 1\unit{TeV}$ is the reference
energy, $\Gamma$ is the spectral index and $E_\mathrm{cut}$ is the
cut-off energy.
The exponential cut-off at high energies will lead to conservative results in terms of the produced 
cascade emission, and will thus lead to conservative constraints on the IGMF.
We construct a likelihood function by combining the likelihood profiles for each \Fermi energy bin\footnote{These are provided in the \Fermi \texttt{SED} file.} with the $\chi^2$ value of the \HESS data points. 
Using only the spectral information (and disregarding the spatial information of the extended halo) can lead, however, to a strong preference for
a dominant cascade contribution.
The reason is that, for a hard intrinsic spectral index and large $E_\mathrm{cut}$,
the combination
of the intrinsic emission and the cascade can perfectly describe the data
(see also the discussion in
Appendix~\ref{app:details_on_the_fit}). However, this scenario would also predict a bright extended halo, which is not observed \citep{HESS:2023zwb}.
Thus, to account partly for the non-observation of a cascade halo, 
we include an additional Gaussian prior on the spectral
index, with a half-width\footnote{This can be deemed as a highly conservative 
    prior, given that the spectral index of \ES vary only by 0.002 
    from $B=10^{-16}$ to 
    $B=10^{-13}\unit{G}$ for $t_\mathrm{max}=10^7\unit{yr}$ in
    \citet{HESS:2023zwb}, where the
    angular information is included in the likelihood.
} $\sigma=0.05$; the effect of the prior is detailed in
Appendix~\ref{app:details_on_the_fit}.
The final likelihood function, i.e. the likelihood of observing the data
$\mathcal{D}_i$ with given model parameters, can be expressed as
\begin{equation}
  \begin{split}
\mathcal{L}(\mathcal{B},\vec{\theta}|\mathcal{D}_i)
=\mathcal{L}(\mathcal{B},\vec{\theta}|\mathcal{D}_{\mathrm{LAT},i})
 \mathcal{L}(\mathcal{B},\vec{\theta}|\mathcal{D}_{\mathrm{\HESS},i}) \\
  \times\frac{1}{\sqrt{2\pi\sigma^2}}
    \exp\left(-\frac{(\Gamma-\Gamma_0)^2}{2\sigma^2}\right),
  \end{split}
\end{equation}
where $\mathcal{B}=(B_0, f, \ldots)$ refers to the parameters of the magnetic
field model and $\vec{\theta}=(N_0, \Gamma, E_\mathrm{cut})$ contains the spectral
parameters.
Meanwhile, we fix the spectral index to the value obtained
in the best fit case, $\Gamma_0=1.64$, which occurs for strong
magnetic fields~(i.e. no cascade halo).
The likelihood fit and the various analyses are discussed in more depth
in Appendix~\ref{app:details_on_the_fit}.

We now proceed to discuss the analyses performed.
First, we set a lower limit on a space-filling cell-like IGMF with magnetic field
strength $B_0$. By virtue of Wilks' theorem, we can convert the likelihood
function to a lower limit
when $2(\ln \mathcal{L} - \ln \mathcal{L}_\mathrm{max})\simeq \chi^2 -
  \chi_\mathrm{min}^2 > \alpha$,
where the number $\alpha$ is the $\chi^2$-quantile.
The subscript `max' refers to the best fit magnetic field strength, which
is achieved for a strong magnetic
field (i.e.~no detectable cascade photons).
Since only one degree of freedom is considered, namely the magnetic
field strength $B_0$, we set a
95\% CL upper limit with $\alpha=2.71$.
We perform a grid search with
$B_0=10^{-17}\unit{G},10^{-17.2}\unit{G}, \ldots, 10^{-12.6}\unit{G}, 
10^{-12.8}\unit{G}$ and use
cubic interpolation to estimate the lower limit.
The final result is that $B_0>5.1\times 10^{-15}\unit{G}$ is excluded
at the 95\% CL for an assumed
blazar activity time of $t_\mathrm{max}=10^{4}\unit{yr}$, while
$B_0>1.0\times 10^{-14}\unit{G}$ for $t_\mathrm{max}=10^{7}\unit{yr}$.
The results for other activity times are detailed
in Appendix~\ref{app:details_on_the_fit}.

Next, we set a lower limit to the filling fraction $f$ of the
top-hat IGMF model (see Fig.~\ref{fig:bfield}).
Since $B_0$ in this case is assumed to be large and the
cascade spectrum is independent of $D\gtrsim 5$ Mpc
(see Appendix~\ref{app:filling}), we are left 
with a single degree of freedom.
As in the previous case, we perform a grid search with
$f=0.00, 0.05,0.10,\ldots,0.95, 1.00$, and use
cubic interpolation to estimate the lower limit. The result is that $f<0.67$ is
excluded with 95\% confidence.

Finally, we compute the likelihood function for 100 line-of-sight magnetic
fields for each of the models discussed in Sec.~\ref{sec:igmf}. In this case,
however, the profile likelihood function cannot be directly applied to exclude
the IGMF models since the exact number of degrees of freedom is unknown 
(we are not varying a single parameter like the magnetic field strength or the filling factor).
The correct approach would be through Monte Carlo simulations, but it is
quite computationally expensive and the
instrumental response function (IRF) of \HESS is not publicly available.
Instead, we note that 
the magnetic field distribution can be approximated by a top-hat
distribution with a filling fraction, $f$, and magnetic field strength, $B$,
whose effects on the likelihood are furthermore correlated.
The exact shape of the cascade spectrum will in general depend on the exact
magnetic field distribution, in particular through the typical size between
structures, $D$, and the correlation length of the magnetic field,
$l_\mathrm{B}$. However, the cascade spectrum remains constant for
$D\gtrsim 5\unit{Mpc}$ (see Appendix~\ref{app:filling})
and $l_\mathrm{B}\gtrsim 100\unit{kpc}$, both of
which are the case in the considered IGMF models.
Thus, we consider a magnetic field model to be excluded by at least 95\%
confidence if $-2(\ln \mathcal{L} - \ln \mathcal{L}_\mathrm{max}) > 4.61$,
corresponding to a $\chi^2$ distribution with two degrees of freedom.
The average and the 95\% quantile of the likelihood estimates for each of the
100 line-of-sights magnetic fields are listed in Tab.~\ref{tab:prop}.
The estimated filling fraction of the models with $|B|>10^{-15}\unit{G}$ is also listed 
(see Appendix~\ref{app:B_properties}).
We rule out a model if the 95\% quantile of the delta log likelihood value is 
above 4.61.
Of the considered MHD simulations without a strong primordial component,
only DYN4 and PCSF survive our test.

\begin{table}
    \begin{tabular}{lccc}
        \hline
         Model  &    $f$ &  $\langle -2\Delta \log\mathcal{L}\rangle$ & 95\%  \\
         \hline
         DYN4   & 0.70 &  3.5  &    0.4  \\
         CSFBH3 & 0.23 & 18.1  &   12.2  \\
         CSF0   & 0.25 & 17.6  &   14.1  \\
         CSF1   & 0.32 & 14.8  &   10.3  \\
         CSF2   & 0.32 & 14.6  &   10.2  \\
         CSF3   & 0.34 & 14.5  &   10.8  \\
         PCSF   & 0.60 &  5.3  &   2.1   \\
        \hline
    \end{tabular}
    \caption{Table containing the estimated filling fraction $f$, mean log likelihood
    estimate and the 95\% quantile of the log likelihood
    estimates for the 100 considered line-of-sights of the MHD simulations.}
    \label{tab:prop}
\end{table}

\section{Discussion and conclusion}

We have used the publicly available data of \ES from \Fermi and
\HESS~\citep{HESS:2023zwb} to set limits on the properties of the IGMF.
By simulating the electromagnetic cascade using \texttt{ELMAG}, we find
that a space-filling IGMF with
$B_0>5.1\times 10^{-15}$ G ($B_0>1.0\times 10^{-14}$ G)
for a blazar activity time of $\Delta t = 10^4$ yr ($\Delta t = 10^7$ yr)
is excluded by 95\% CL.
Note, however,  that the source has only been observed for $\sim 20$ years,
which entails a lower limit on the activity time. In the conservative case $\Delta t=20$ yr,
we obtain
$B_0>2.4\times 10^{-16}\unit{G}$.
Thanks to our prior on the spectral index, we are able to 
account (partially) for the spatial information of \Fermi and \HESS. Thus, our limits
are among the most stringent limits on the IGMF,
only slightly weaker than those obtained with full spatial
information~\citep{HESS:2023zwb}. Note that the fact that our results
are perfectly compatible to what was found in the \HESS analysis
by \citet{HESS:2023zwb} can be considered as an additional motivation
for the prior.

The focus in this work has been on constraining an astrophysical origin
of the IGMF. We have set new limits on the filling fraction $f>0.67$
using a top-hat distribution of the magnetic field.
Note that our result can be considered as conservative
since we assume a strong magnetic field in the filaments:
Decreasing $B_0$ would lead to a greater cascade spectrum, implying that
a larger filling $f$ is needed to avoid over-producing the observed spectrum.
It is worth mentioning that this result is similar to the results obtained by~\citet{Dolag:2010ni} based
on old upper limits of the spectrum of \ES from \Fermi. However, the limits obtained in this
work are statistically more robust since the spectral index, cut-off energy and
normalization are included as nuisance parameters.
The limit can in principle be used to get approximate bounds on the more
realistic magnetic field models that we have considered in this
work. Indeed, from the estimated filling fraction
of the MHD models in Tab.~\ref{tab:prop}, one can foresee that all purely astrophysical models we examined
(CSFBH3, CSF0, CSF2 and CSF3) can be excluded with high significance, while DYN4 survives
the likelihood test. Meanwhile, PCSF may survive the test due to cosmic
variance, but will be in tension with the data.

We performed a likelihood analysis of a variety of simulated
magnetogenesis scenarios.
Due to the cosmic variance in the MHD simulations, the properties of the
magnetic field (e.g.,~effective filling fraction)
depend on the exact line of sight considered in the
simulation volume. Therefore, we conducted the likelihood test on 100
random line-of-sights for each model, and
consider the model excluded at the 95\% CL if its 95\% quantile has
a log likelihood value $-2\Delta \log \mathcal{L}>4.61$, corresponding to a
95\% exclusion of a $\chi^2$ distribution with two degrees of freedom.
Only the models DYN4 and PCSF cannot be excluded at the 95\% CL.
Both may likely be excluded in an improved analysis including more sources
and complete spatial information from \HESS, which currently is not
publicly available.
However, DYN4 and PCSF can be rejected as viable possibilities
based on other complementary tests: The high efficiency dynamo amplification
model DYN4 has been shown to yield a too large amount of
Faraday rotation for polarized radio sources shining through the
cosmic web, at variance with real LOFAR
observations~\citep[e.g.,][]{OSullivan:2020pll,Vazza:2021vwy}.
The PCSF model instead employs by construction a too large feedback
efficiency for low mass halos, yielding inconsistent scaling relations
for halos and star formation histories~\citep{Vazza:2017qge} and
it has been employed here just to gauge the maximum conceivable
contribution from astrophysical seeding. 

A few caveats of our results
should be mentioned (see also the discussions in \citet{HESS:2023zwb}).
Perhaps most importantly, the limits depend strongly on the assumed blazar activity time,
reducing by about an order of magnitude for every second decade in activity time;
the detailed dependence is given in Appendix~\ref{app:details_on_the_fit}.
Moreover, it has been argued that the electrons and positrons may predominantly lose their
energy through plasma instabilities if the blazar activity time is
$\Delta t\gtrsim \mathcal{O}(10^2)\unit{yr}$~\citep{Broderick:2011av}. If confirmed,
it will invalidate the limits based on cascade emissions from TeV blazars. Safe to say,
however, the effect is still debated~\citep{Rafighi:2017ise,AlvesBatista:2021sln}.

In conclusion, with our work we have significantly  narrowed down the possibility that the
IGMF can be of pure astrophysical origin, because the necessary fraction of the space which
must be filled by a strong magnetic field appears impossible to be obtained by all astrophysical 
scenarios for the magnetization of the cosmic web we tested. The upcoming Cherenkov Telescope
Array (CTA) will improve the constraints further~\citep{Meyer:2016xvq,CTA:2020hii}.

\begin{acknowledgments}
The authors would like to thank M.~Kachelrie\ss{} for fruitful discussions and for providing valuable comments on the manuscript. J.T.\ would like to express gratitude for the hospitality at the University of Agder (UiA).  
M.M.\ acknowledges the support from the Deutsche Forschungsgemeinschaft (DFG, German Research Foundation) under Germany’s Excellence Strategy -- EXC 2121 ``Quantum Universe'' -- 390833306 and from the European Research Council (ERC) under the European Union’s Horizon 2020 research and innovation program Grant agreement No. 948689 (AxionDM).
F.V.\ acknowledges financial support from  the Cariplo ``BREAKTHRU'' funds Rif: 2022-2088 CUP J33C22004310003.
In this work, we used the {\enzo} code (\url{http://enzo-project.org}), the product of a collaborative effort of scientists at many universities and national laboratories. Their commitment to open science has helped make this work
possible. F.V. gratefully acknowledges the Swiss National Supercomputing Centre (CSCS), Switzerland, for the access to the Piz Daint, under the allotted project ``s1096''.  
\end{acknowledgments}


\newpage

\appendix

\section{Details of the magnetic field models}
\label{app:B_details}

\subsection{Cosmological MHD simulations}
\label{app:MHD}

In this work, we make use of a selection of models from a suite of MHD runs employing a fixed mesh resolution, known as the ``Chronos++ suite'', which encompasses numerous models exploring various plausible scenarios for the evolution of large-scale magnetic fields. 
More specifically, we use a  ``baseline'' non-radiative simulation with a simple uniform primordial seeding of magnetic fields, which we contrast to additional 
re-simulations with radiative gas cooling and different variations of star and AGN feedback, as well as a non-radiative model in which a simple sub-grid model was used to estimate the maximum conceivable small-scale dynamo amplification.  
The models including radiative gas, star formation and feedback from supermassive black holes (SMBHs) were selected from a larger suite of simulations, with the prior that their predicted cosmic star formation history and scaling laws on simulated halos (e.g. mass--temperature relation) provide a reasonable match with observations, as discussed by \citet{Vazza:2017qge}. 
Although most of the models have been shown to produce predictions incompatible with the latest results from radio surveys~\citep{Vazza:2021vwy}, we agnostically test them against blazar spectra to check whether additional and independent constraints could be derived.  

All the runs adopt the $\Lambda$CDM cosmology, with density parameters $\Omega_{\rm BM} = 0.0478$ (BM representing the Baryonic Matter), $\Omega_{\rm DM} = 0.2602$ (DM being the Dark Matter)  and $\Omega_{\Lambda} = 0.692$ ($\Lambda$ being the cosmological constant), and a Hubble constant $H_0 = 67.8$ km/sec/Mpc \citep{Planck:2015fie}.  The initial redshift is $z=38$, the spatial  resolution is $83.3 ~\rm kpc/cell$ (comoving) and the constant mass resolution of the dark matter particles is set to $m_{\rm DM}=6.19 \cdot 10^{7}M_{\odot}$. 

While we refer the reader to \citet{Vazza:2017qge} for more detailed explanations on the physical and numerical parameters of these simulations, we give here a short summary of the most important ingredients, whose variations are shown in Tab.~\ref{tab:mhd}:

\begin{table}[bht]
    \centering
    \begin{tabular}{ccccccclcc}
         \hline
         ID &  cooling?&  SF?&  $n_{SF}$&  $t_{SF}$&  $\epsilon_{SF}$&  $\epsilon_{B,SF}$&   SMBH?&$B_0$&  description\\
 & & & $[\rm part/cm^3]$& $[\rm Gyr]$& & &  &$[\rm G]$& \\ \hline 
         P&  n&  n&  -&  -&  -&  -&   n&$10^{-9}$&  high $B_0$\\
         DYN4&  n&  n&  -&  -&  -&  -&   n&$10^{-18}$&  weak $B_0$, dynamo\\
 CS& y& y& $10^{-3} $& $1.5\rm$& 0& 0&  n&$10^{-9}$& no stellar fb,\\
 CSF0& y& y& $10^{-3} $& $1.5\rm$& $10^{-9}$& 0.01&  n&$10^{-18}$&  very low stellar fb.\\
         CSF1&  y&  y&  $10^{-3} $&  $1.5\rm$&  $10^{-8}$&  0.01&   n&$10^{-18}$&  low stellar fb.\\
         CSF2&  y&  y&  $5 \cdot 10^{-4}$&  $1.5\rm$&  $10^{-7}$&  0.1&   n&$10^{-18}$&  medium star fb.\\
         CSF3&  y&  y&  $5 \cdot 10^{-4}$&  $1.0\rm $&  $10^{-6}$&  0.1&   n&$10^{-18}$&  high star feedback\\
         CSFDYN1&  y&  y&  $2 \cdot 10^{-4} $&  $1.0\rm $&  $10^{-6}$&  0.1&   n&$10^{-9}$&  high $B_0$, high star fb., dynamo\\ 
         CSF5&  y&  y&  $2 \cdot 10^{-4}$&  $1.0\rm $&  $10^{-6}$&  0.1&   n&$10^{-11}$&  low $B_0$, high star fb.\\ 
         PCSF&  y&  n&  0&  0&  0&  0.01&   y, fixed E.&$10^{-18}$& very efficient SMBH fb.\\
         CSFBH3&  y&  y&  $5 \cdot 10^{-4}$&  $1.5\rm $&  $10^{-8}$&  0.01&   y, Bondi&$10^{-18}$&  low star fb,  SMBH fb.\\ 
         \hline
    \end{tabular}
    \caption{Summary of the various simulation ingredients and the differences between the
    magnetogenesis scenarios analysed in this work. The meaning of the columns is as follows: (ID) model name, (cooling?) presence of radiative gas cooling, (SF?) presence of star formation, ($n_\mathrm{SF}$) threshold gas density for star formation, ($t_{SF}$) fixed timescale for star formation, ($\epsilon_{SF}$) efficiency of star feedback, ($\epsilon_{B,SF}$) fraction of feedback energy delivered into magnetic energy, (SMBH?) presence of supermassive black holes, ($B_0$) initial primordial magnetic field in the simulation, (description) additional description of the run.}
    \label{tab:mhd}
\end{table}

\begin{itemize}
    \item{\it initial magnetic field ($B_0$)}:
    the magnetogenesis scenarios assume the existence of a weak and volume-filling magnetic field at the beginning of the simulation. These are the only seed of primordial magnetization in the simulated Universe. A spatially uniform seed field value is set for every component of the initial magnetic field, using in this subset of runs $B_0=10^{-9}$, $10^{-11}$ or $10^{-18}$ G (comoving).
    \item {\it sub-grid dynamo}:
    the small-scale dynamo amplification of magnetic fields in the simulation is computed at run-time via sub-grid modelling, based on the estimated rate of dissipation of solenoidal turbulence. These models attempt to bracket the possible residual amount of dynamo amplification which can be attained, on energy grounds, in the cosmic structures, but may be lost because of a lack of resolution and/or for the onset of small-scale plasma instabilities that cannot be captured from first principles in the simple hydro/MHD model. 
    \item {\it radiative cooling}:
    standard equilibrium gas cooling implemented in ENZO, based on the chemical distribution of 7 main elements (HI, HII, HeI, HeII, HeIII, free electrons and "metals").
    \item {\it star formation}: 
    star formation criterion~\citep{Kravtsov:2003gx} included in {\enzo}.
    Here, new star particles are formed on the fly if the local gas density exceeds a given threshold, $n_{*}$, with a dynamical timescale, $t_{*}$. The minimum star mass in the considered simulations is set to $m_{*}=10^7 M_{\odot}$. Whenever $(i)$ $n \geq n_{*}$, $(ii)$ the gas is contracting ($\nabla \cdot \vec{v} < 0$), $(iii)$ the local cooling time is smaller than the dynamical timescale ($t_{\rm cool} \leq t_*$), and $(iv)$ the baryonic mass is larger than the minimum star mass ($m_b=\rho \Delta x^3 \geq m_*$), stars with a mass $m_* =m_b \Delta t/t_{\rm *}$ are set to form.  This recipe was introduced to reproduce the observed Kennicutt's law~\citep{Kennicutt:1997ng}.
    \item {\it feedback from star formation}:
    stars return thermal energy, gas and metals back into the gas phase. The feedback from the star forming process (e.g. supernova explosions) depends on the assumed fractions of energy/momentum/mass ejected for each formed star particle, $E_{SN}= \epsilon_{SF} m_* c^2$. The feedback energy is assumed to be released entirely in thermal form (i.e. hot supernovae-driven winds), which typically turns into the formation of  pressure-driven winds around the halos ($v_{\rm wind} \sim 10-10^2$ km/s).  
    For an overview of the performances and limitations of the recipes for star formation and feedback in {\enzo} we refer the reader to the recent survey of models by \citet{Skory:2012ni} and \citet{Hlavacek-Larrondo:2022frz}.  The limited  spatial and mass resolution is expected to quench the formation of $ \leq 10^8 M_{\odot}$ galaxies, implying an important contributor to the cosmic star formation rate \citep[e.g.][]{Genel:2014lma}.  However,  the contribution to the chemical, thermal and magnetic enrichment of the intergalactic medium by galactic outflows is predicted to be dominated by $\sim 10^{10}-10^{11} M_{\odot}$ \citep[e.g.][]{Samui:2017dsz} galaxies, which are formed in the runs considered here.
    \item {\it supermassive black holes (SMBHs)}:
    black hole (BH) seed particles are injected at $z=4$ at the centre of all $\geq 10^{9} M_{\odot}$ halos identified in the simulation~\citep{Kim:2011ub}. For each seed particle, a fixed initial mass of $M_{\rm BH}=10^4 M_{\odot}$ is assumed.   Each BH particle accretes gas mass according to the Bondi--Hoyle rate, assuming all gas gets accreted at the fixed temperature of $3 \times 10^5 \unit{K}$ (because a precise estimate of gas temperature on the accretion radius is not feasible with the given resolution) and introducing a numerical boost to the measured accretion rate by $\alpha_\mathrm{Bondi}=100$ to overcome the effect of the limited gas density which can be resolved around the BH particles. 
    \item {\it supermassive black hole (SMBH) feedback}:
    as an extra thermal energy output from each BH particle, they release thermal feedback on the surrounding gas, assuming an efficiency $\epsilon_{\rm BH} = \Delta M c^2 \Delta t/E_{\rm BH}$  between the accreted mass, $\Delta M$,  and the feedback energy, $E_{\rm BH}$. In the PCSF model variation, a ``maximal'' SMBH feedback scheme is used instead, in which every SMBH is forced to release the same fixed amount of feedback thermal energy,  $E_{\rm BH}=5 \times 10^{59} \rm erg$, regardless of the mass accretion rate. This model is extreme, in the sense that it overpredicts the typical feedback energy in the lowest mass galaxies, up to $\sim 100\text{--}1000$ factors in power; however, it is meant to bracket the maximum possible effect of astrophysical magnetization in the simulation volume. 
    \item {\it magnetic field seeding}:
    For the injection of magnetic fields by stellar feedback, the generation of magnetic dipoles during each feedback episode is introduced  in {\enzo}. The dipoles are randomly oriented along one of the coordinate axes of the simulation, and their total energy scales with the thermal feedback energy as $E_{\rm b,SN}=\epsilon_{\rm SF,b} \cdot E_{\rm SN}$, typically with $\epsilon_{\rm SF,b}~\sim 1\text{--}10 \%$.  Also for the release of magnetic energy from SMBH, magnetic fields from the BH region are injected (in the form of magnetic dipoles) assuming $E_{\rm b,AGN}= \epsilon_{\rm b,AGN} E_{\rm BH}$, where values in the range $\epsilon_{\rm b,AGN}=1\text{--}10 \%$ of the energy conversion into magnetic field dipoles at the jet base were explored. 
\end{itemize}

While the fixed spatial resolution of the runs is not good enough to model properly the galaxy formation processes with sufficient detail (even massive galaxies are resolved within a few cells at most), it was shown by \citet{Vazza:2017qge} how the star-formation recipes are specifically calibrated to reproduce properly the cosmic star formation history in the volume. Moreover, the combination of cooling and stellar/AGN feedback has been tuned so that the scaling relations of galaxy clusters and groups are well reproduced. The models also describe the innermost density, temperature, and entropy profiles of galaxy clusters fairly well in comparison to higher resolution simulations. In summary, previous work suggest that the feedback recipes implemented in the considered simulations can effectively mimic the large-scale effect and energetics of galaxy formation recipes at a higher resolution, hence they can be effectively used to model the large-scale effects of thermal and magnetic feedback on the surrounding distribution of filaments.

\subsection{Properties of the magnetic fields}
\label{app:B_properties}

\begin{figure}[htb]
    \centering
    \includegraphics[scale=\figscale]{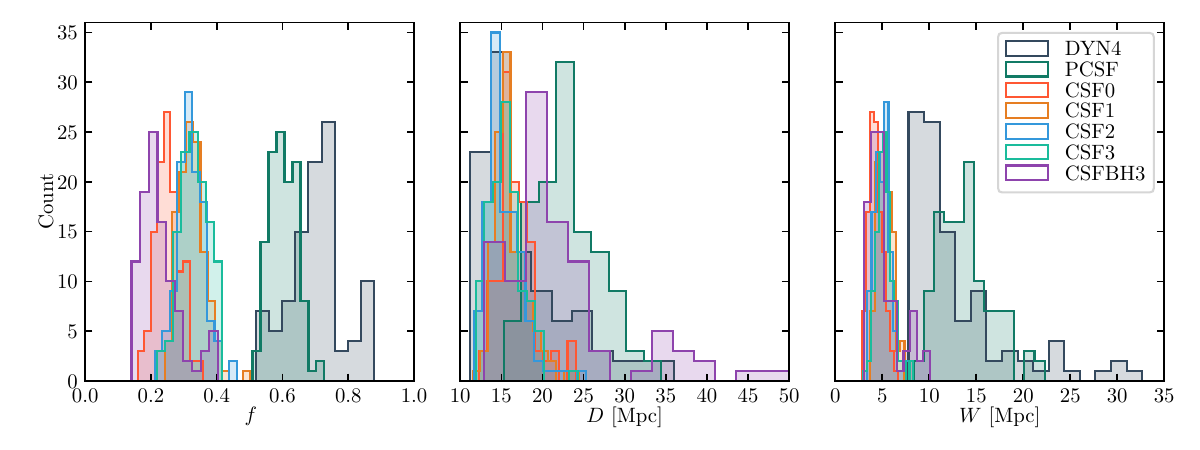}
    \caption{Distributions of the degree of filling, $f$, distance
    between filaments, $D$, and size of filaments $W$ of the
    100 line-of-sights that are considered in the likelihood
    analysis of the MHD models. The different colours correspond to
    different models.}
    \label{fig:DfW}
\end{figure}

In a typical IGMF of astrophysical origin, the propagation of
the electrons and positrons in an electromagnetic cascade can be split into
two distinct regimes:
When the electron Larmor radius, $r_\mathrm{L}$, is much smaller than the
inverse Compton scattering length, $\lambda_\mathrm{IC}$, they move
ballistically. This is the case in the voids, where the magnetic field
is weak or potentially non-existent. On the other hand, when
$r_\mathrm{L}\gg \lambda_\mathrm{IC}$, they will be isotropized and
contribute to the diffuse gamma-ray background. This is
typically the case in the filaments.
The typical transition between the two
regimes is where the Larmor radius of the electrons is of the same order
as the inverse Compton scattering length.
That is, $r_\mathrm{L}\sim \lambda_\mathrm{IC}$, which in turn leads to
$E_e/\mathrm{TeV}\sim B/\mathrm{10^{-15}\unit{G}}$.
In order to get a more intuitive understanding of the magnetic field models and their
effect on the cascade spectrum, we compare them with a simple top-hat distribution of
the line-of-sight magnetic field, wherein the filaments have a fixed size and
mutual distance. Furthermore, it is assumed that
$B\sim 10^{-9}\unit{G}$ in the filaments,
so that $r_\mathrm{L}\ll\lambda_\mathrm{IC}$. Thus, the model can be characterized by
two numbers: the distance between filaments, $D$, and the filling fraction, $f$. We extract the
numbers from the line-of-sights by defining filaments as regions with magnetic field 
strength larger than $10^{-15}\unit{G}$. Furthermore, we demand that a void has
to be larger than $333\unit{kpc}$ (i.e., four times the simulation grid). The distributions of $D$ and $f$ for the considered
IGMF models are summarized in Tab.~\ref{tab:DfW} and Fig.~\ref{fig:DfW}.

\begin{table}[htb]
\centering
\begin{tabular}{lccc}
\hline
    Model    & $f$    & $D$ [Mpc] & $W$ [Mpc] \\ \hline
    P        &  1.00  &     --    &     --    \\
    DYN4     &  0.70  &   12.5    &    8.8    \\
    CS       &  1.00  &     --    &     --    \\
    CSF0     &  0.25  &   15.7    &    4.0    \\
    CSF1     &  0.32  &   15.3    &    4.9    \\
    CSF2     &  0.32  &   14.8    &    4.7    \\
    CSF3     &  0.34  &   14.6    &    5.0    \\
    CSFDYN1  &  1.00  &     --    &     --    \\
    CSF5     &  1.00  &     --    &     --    \\
    PCSF     &  0.60  &   23.6    &   14.3    \\
    CSFBH3   &  0.23  &   21.1    &    4.5    \\
\hline
\end{tabular}
\caption{Summary of the average effective
    filling fraction, $f$, distance between
    magnetic field structures, $D$, and width of magnetic field structures,
    $W$ (cf.~Fig.~\ref{fig:bfield}) of the considered
    MHD models listed in Tab.~\ref{tab:mhd}.
}
\label{tab:DfW}
\end{table}

\section{Details on the cascade simulations with \texttt{ELMAG}}
\label{app:simulation}

For each magnetic field model, we inject $10^5$--$10^7$ photons\footnote{
  We used $10^5$ photons for each of the 100 line-of-sight magnetic fields
  for each of the considered MHD models, $10^7$ photons for the space-filling
  domain-like field, and $10^6$ photons in all other cases.
} logarithmically
randomly distributed between 1 GeV and 100 TeV. The results are stored as
a four-dimensional array depending on the time-delay, $\Delta t$, the
primary energy of the injected photon, $E_i$, the energy of the arriving photon,
$E_f$, and the arrival angle, $\theta$.
For the energies, 10 bins per decade are used. 
Meanwhile, for the time-delay, one bin per decade
between $\Delta t = 1$ and $10^7$ years is used, in addition to $\Delta t < 1$ and
$\Delta t > 10^7$ years. 
Finally, for the arrival direction, 500 bins are linearly distributed between
$\theta=0^\circ$ and $1.5^\circ$. 

Note that some changes to \texttt{ELMAG 3.02} were needed for this work
\citep[see also][]{Kachelriess:2021rzc,Kachelriess:2023fta}:
\begin{itemize}
    \item Inclusion of a domain-like magnetic field
    \item Inclusion of line-of-sight magnetic fields provided by a file; The field strength is set to
    the projection of the magnetic field along the line of sight
    \item Export of results as multidimensional histograms
    \item Inclusion of a top-hat magnetic field
    \item Fine-tuning the ordinary differential equation solver,
    such that one can consider magnetic fields stronger than $\sim 10^{-13}\unit{G}$
\end{itemize}

\section{Details on the fit}
\label{app:details_on_the_fit}

\begin{figure}[htb]
    \centering
    \includegraphics[scale=\figscale]{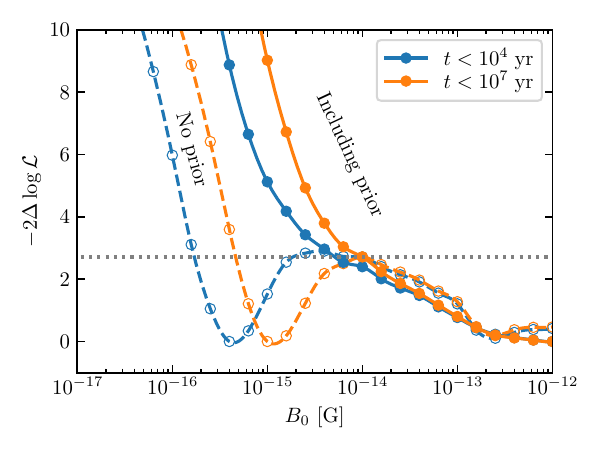}
    \includegraphics[scale=\figscale]{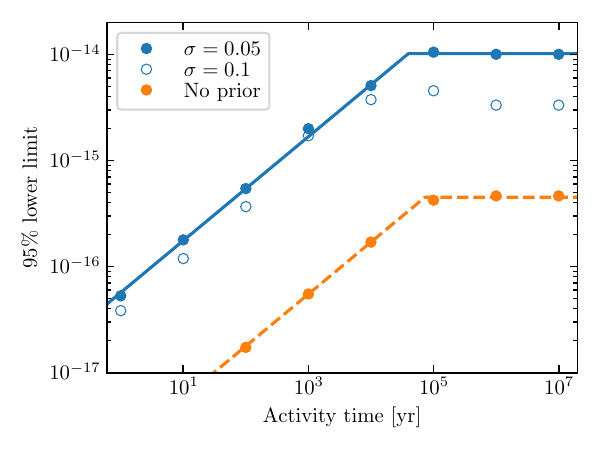}
    \caption{The likelihood profile obtained for a domain-like, space-filling
    IGMF as a function of the magnetic field strength $B_0$ is plotted (left).
    The result is shown both with (dashed lines) and without (solid lines) the
    Gaussian prior on the spectral index, as well as blazar activity times
    $t=10^4$ yr (blue) and $t=10^7$ yr (orange). Magnetic fields giving a
    likelihood value above the gray dotted line are excluded. The dependence
    of the limit on the activity time is plotted to the right.
    }
    \label{fig:loglike}
\end{figure}

We construct the log-likelihood estimate using the publicly available spectral
information of the blazar \ES from \Fermi and \HESS~\citep{HESS:2023zwb},
and minimize it with a spectral
reweighting~\citep{Fermi-LAT:2018jdy} of the multidimensional 
histogram discussed in Appendix~\ref{app:simulation} using
using the
\texttt{differential\_evolution} algorithm in \texttt{scipy.optimize}.
The resulting likelihood profile is plotted with dashed lines
in the left panel of Fig.~\ref{fig:loglike} for blazar activity times $10^{4}$ yr (blue) and
$10^7$ yr (orange) for a domain-like magnetic field.
The dependence of the lower limit on the activity time is shown in the right
panel of Fig.~\ref{fig:loglike}. For comparison, we also plot the limits with a wider prior
$\sigma=0.1$ (open circles).
The limits as a function of the activity times
are well
reproduced\footnote{This fit function can be motivated by the observation that 
the time delay scales with the IGMF as
$t\propto B^2$~\citep[see Eq.~(19) in][]{AlvesBatista:2021sln}, and should become constant
since we consider the 95\% containment region around the source.
}
as $B_0 \sqrt{\Delta t/\mathrm{yr}}$, becoming
constant at $t_0$. The best fits\footnote{We exclude $\Delta t=10^{4}$ and $10^{5}\unit{yr}$
from the fit.}, shown in Fig.~\ref{fig:loglike}, are
$\{t_0=3.38\times 10^{4}\unit{yr},~B_0=5.46\times 10^{-17}\unit{G}\}$ with the prior on the
spectral index ($\sigma=0.05$), and
$\{t_0=7.08\times 10^4\unit{yr},~B_0=1.74\times 10^{-18}\unit{G}\}$ without the prior.

We have tested that the results are consistent with those obtained using a 
Gaussian turbulent field, a rescaling of the pure primordial magnetogenesis scenario (P),
and the 2.5D implementation in \texttt{ELMAG}. 

\begin{figure}[htb]
    \centering
    \includegraphics[scale=\figscale]{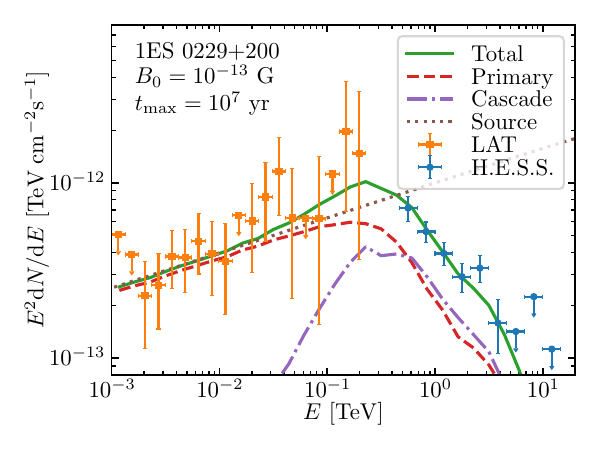}
    \includegraphics[scale=\figscale]{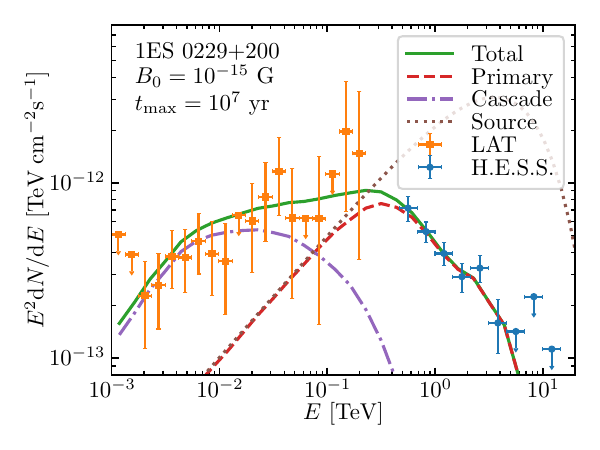}
    \includegraphics[scale=\figscale]{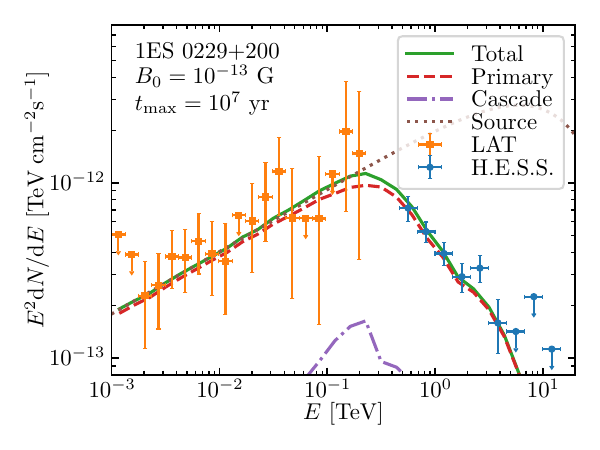}
    \includegraphics[scale=\figscale]{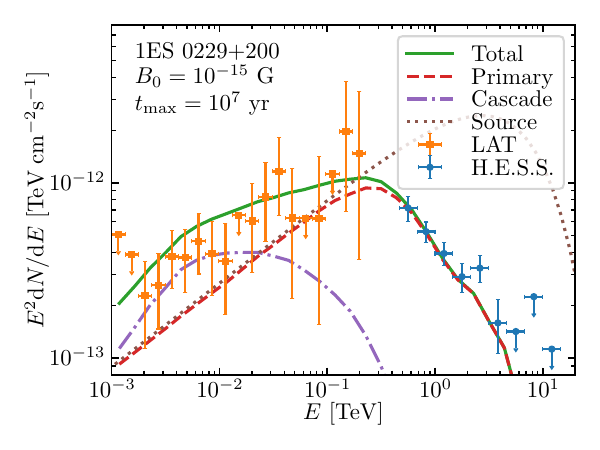}
    \caption{
        Visualization of the contributions to the measured spectrum
        (orange error bars for \Fermi, blue for \HESS) of \ES:
        The source spectrum (brown dotted line) is
        attenuated by the EBL, so that 
        an attenuated primary spectrum (red dashed line) is observed on
        Earth. In addition, the electromagnetic cascade may induce
        a spectrum of cascade photons (purple dotted dashed line)
        at a reduced energy. The source spectrum is adjusted such that
        the total spectrum (green solid line) minimizes the likelihood
        estimate of the measured data.
        The results are shown for $B_0=10^{-13}\unit{G}$ (left column)
        and $B_0=10^{-15}\unit{G}$ (left column), as well as
        without the Gaussian prior on the spectral index
        (top row) and with the Gaussian prior (bottom row).
    }
    \label{fig:spectrum_domain}
\end{figure}

In the likelihood profile (left panel in Fig.~\ref{fig:loglike}),
it is apparent that there is an allowed region
around $B_0\sim 10^{-15}\unit{G}$, which is due to a strong preference for a
large cascade contribution in the fit, i.e. large cut-off energy and small spectral index.
This is well visualized in the spectra plotted in Fig.~\ref{fig:spectrum_domain}.
However, a significant cascade contribution has already been excluded due to the
non-observation of the gamma-ray halo~\citep{HESS:2023zwb}.
To accommodate for this effect, we include a Gaussian prior on the
spectral index with half-width $\sigma=0.05$,
a conservative prior
given that the spectral index only vary by 0.002 for $\Delta t=10^7\unit{yr}$ in~\citet{HESS:2023zwb}.
In this case, the minimum vanishes, and the results are on par with those
of~\citet{HESS:2023zwb}. The results are shown in Fig.~\ref{fig:loglike} in solid
lines.

\subsection{Limit on the filling fraction}
\label{app:filling}

\begin{figure}[htb]
\centering
\includegraphics[scale=\figscale]{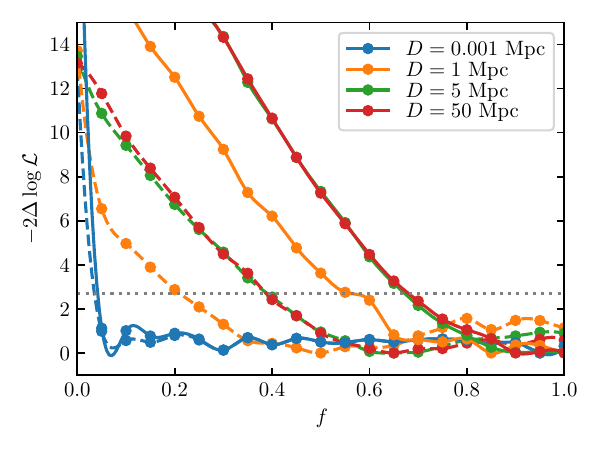}
\caption{Likelihood profile for the filling fraction for the top-hat distributed
    magnetic field. The result is shown both with (solid lines) and without
    (dashed lines) the prior on the spectral index, as well as for
    various values of the size of the magnetic field clusters, $D$.
    Values of $f$ giving a
    likelihood value above the gray dotted line are excluded.
}
\label{fig:loglike_top_hat}
\end{figure}

The likelihood profile for a simple top-hat distribution as a function of the
filling fraction is shown in Fig.~\ref{fig:loglike_top_hat} for various distances
between structures, $D$.
It is clear that the results are independent of $D$ for $D\gtrsim 5\unit{Mpc}$,
which is the case for the magnetic field models considered in this work. 
As discussed in the
previous subsection, a preference for a significant cascade halo in the fit extends the
allowed filling fraction down to $f>0.39$. However, when including the bound on the 
cascade halo, the limit becomes $f>0.67$.

\subsection{Limits on the MHD magnetic fields}

\begin{figure}[htb]
\centering
\includegraphics[scale=\figscale]{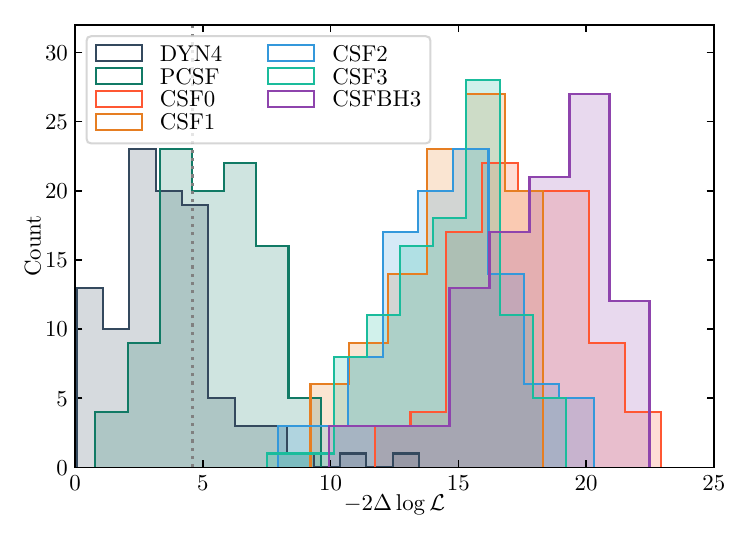}
\caption{
  Histogram of the likelihood estimate for the various line-of-sights
  of the MHD models. Line-of-sights to the right of the dashed grey line
  are excluded.
}
\label{fig:loglike_mhd}
\end{figure}

\begin{figure*}[htb]
\centering
\includegraphics[scale=\figscale]{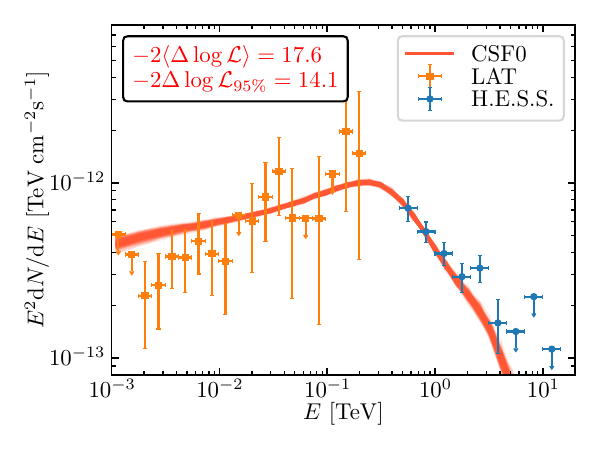}
\includegraphics[scale=\figscale]{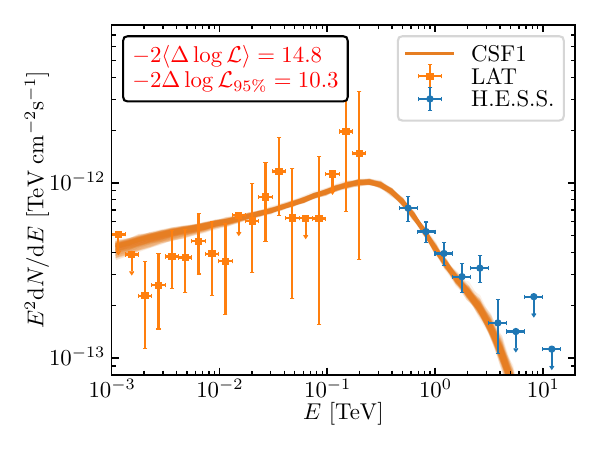}
\includegraphics[scale=\figscale]{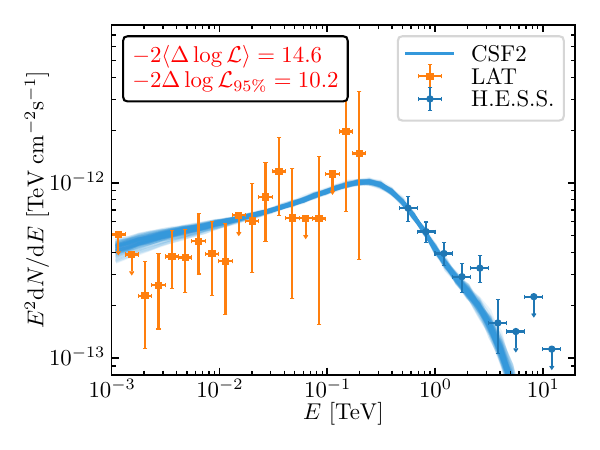}
\includegraphics[scale=\figscale]{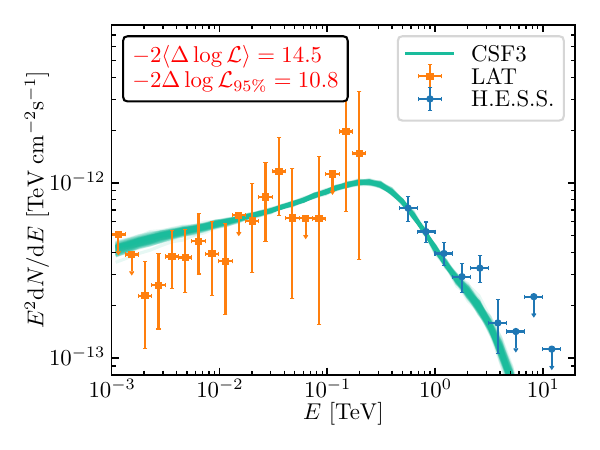}
\includegraphics[scale=\figscale]{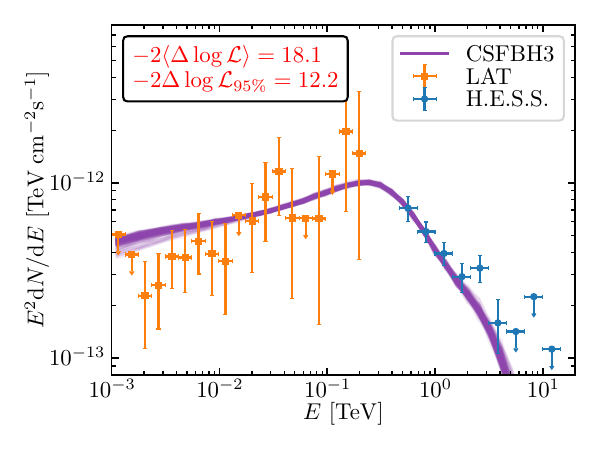}
\caption{
  Same as Fig.~\ref{fig:cascade_example} for the rest of
  the considered MHD models.
}
\label{fig:spectrum_mhd}
\end{figure*}

Here, we include some additional plots for the analysis of the MHD
models. In Fig.~\ref{fig:loglike_mhd}, we plot the histogram of the
likelihood estimate for the various line-of-sights of the MHD models.
Meanwhile, in Figs.~\ref{fig:cascade_example} and~\ref{fig:spectrum_mhd}, we plot
the corresponding best-fit spectra for all the line-of-sights.

\end{document}